\documentclass[aps,prl,twocolumn]{revtex4-1}
\usepackage[english]{babel}
\usepackage{amsmath}
\usepackage{graphicx}
\usepackage{comment}
\usepackage{xcolor}

\usepackage{ulem}

\begin{document}

\title{Effect of particle shape on the flow of an hourglass}

\author{Bo Fan,\textit{$^{1,3}$},Tivadar Pong\'o,\textit{$^{2,4,1}$}
Ra\'ul Cruz Hidalgo,\textit{$^{2}$} and Tam\'as B\"orzs\"onyi\textit{$^{1}$} 
} 

\address{
$^1$Institute for Solid State Physics and Optics, HUN-REN Wigner Research Centre for
Physics, P.O. Box 49, H-1525 Budapest, Hungary\\
$^2$F\'isica y Matem\'atica Aplicada, Facultad de  Ciencias, Universidad de Navarra, Pamplona, Spain\\
$^3$Physical Chemistry and Soft Matter, Wageningen University $\&$ Research, Wageningen, The Netherlands \\
$^4$Collective Dynamics Lab, Division of Natural and Applied Sciences, Duke Kunshan University, 215306, Kunshan, Jiangsu, China}

\vspace{10pt}

\begin{abstract}
The flow rate of a granulate out of a cylindrical container is studied as a function of particle shape 
for flat and elongated ellipsoids experimentally and numerically. We find a nonmonotonic dependence of the flow rate on the grain aspect ratio $a/b$. Starting from spheres the flow rate grows and has two maxima around the aspect ratios of $a/b\approx 0.6$ (lentil-like ellipsoids) and $a/b\approx 1.5$ (rice-like ellipsoids) reaching a flow rate increase of about $15\%$ for lentils compared to spheres. For even more anisometric shapes ($a/b=0.25$ and $a/b=4$) the flow rate drops. Our results reveal two contributing factors to the nonmonotonic nature of the flow rate: both the packing fraction and the particle velocity through the orifice are nonmonotonic functions of the grain shape. Thus, particles with slightly non-spherical shapes not only form a better packing in the silo but also move faster through the orifice than spheres. We also show that the resistance of the granulate against shearing increases with aspect ratio for both elongated and flat particles, thus change in the effective friction of the granulate due to changing particle shape does not coincide with the trend in the flow rate. 
\end{abstract}

%
%
%
%
%
\maketitle

Flow of a granular material out of a container is a common process in everyday life,  agriculture and industrial operations. 
Typically, a granulate discharges through the orifice with a constant flow rate, independent of the filling height \cite{fowlerCES1959,beverlooCES1961}.  This feature was used when the hourglass was constructed as a time measuring device long ago, and it is very useful as the required flow rate can be easily set by simply choosing the appropriate orifice size.  The flow rate changes with increasing orifice size as a power law function (Beverloo law \cite{beverlooCES1961,neddermanCES1982,mankocGM2007,hiltonPRE2011,rubiolargoPRL2015,calderonPT2017}) and also depends on the internal friction of the granular material, which is changing with the surface roughness as well as the  shape of the grains. 
Naturally, increasing particle roughness negatively impacts the flow rate, but it is much less obvious how it should change with grain shape.
On one end of the spectrum, very irregular grains can get entangled during the discharge and flow less easily, but what should we expect from shapes which deviate only slightly from a sphere: ellipsoids with rice-like or lentil-like shapes?

Elongated or flat particles are observed to develop orientational ordering in a {\it shear flow}, with their smallest cross section facing almost in the flow direction \cite{borzsonyiPRL2012,borzsonyiPRE2012,borzsonyiSM2013,borzsonyiNJP2016,botonPRE2013}. The average orientation angle decreases with increasing grain anisometry, and e.g. is around $10^\circ$ for a rice-like ellipsoid with elongation $a/b=3$. Naively, this would then suggest easier flow and thus faster flow rate through a constriction for simple elongated or flat ellipsoids than for spherical particles. However, taking a closer look at the dynamics of such particles in a shear flow, we observe that they perform irregular rotation as dictated by the shear stress related to the interaction with their neighbors.
On average, they rotate slower when they are nearly parallel to the flow direction and rotate faster when they are perpendicular to it \cite{borzsonyiPRE2012}. So during their rotation, they spend most of the time
nearly parallel to the flow, which leads to the above-described average orientation. But as they rotate, neighboring particles actually get into conflict and hinder each other’s motion. This leads to a nontrivial rheology for such types of granular materials.

Previous studies focusing on the fundamental question of the effect of the grain shape on the {\it rheology} of a granular material or {\it discharge rate from a silo} are mostly {\it numerical}. This is because discrete element model (DEM) simulations offer a straightforward way to systematically change the particle shape without changing the other parameters (microscopic surface friction, etc.). As for the {\it rheology}, recent DEM studies show that for frictional particles the effective friction $\mu_\text{eff}$ of the system is increasing with grain anisometry. This was found for spherocylinders in quasistatic shear flow for interparticle friction $\mu_\text{p}>0.4$ \cite{nagyNJP2020}, or in more dynamic inclined plane flows for $\mu_\text{p}\geq0.5$ \cite{hidalgoPRF2018}. Similar observations were made in a simplified two-dimensional (2D) system \cite{trulssonJFM2018}. Interestingly, for systems with lower interparticle friction ($\mu_\text{p}<0.4$) a nonmonotonic tendency was found: starting from a spherical shape $\mu_\text{eff}$ first increases and then decreases with  $a/b$.
Focusing on previous DEM results on the {\it discharge of a 3D silo} with frictional grains one finds contradictory observations. 
On one hand, Liu et al.~found a reduced discharge rate for both elongated and flat ellipsoids compared to the case of spheres \cite{liuPT2014}, on the other hand, Li et al.~reported a larger flow rate for round disks than for spheres \cite{liCES2004}. 
In a recent work by Hesse et al.~decreased/increased flow rate was found for elongated/flat ellipsoids compared to spheres, respectively \cite{hessePT2021}. Finally, for frictionless grains Langston et al.~reported the same flow rate for spherocylinders and spheres \cite{langstonCES2004}.

To our best knowledge, so far no systematic {\it experimental} tests have been performed to measure how the discharge rate changes with particle shape when all other parameters (surface roughness, density, hardness, etc.) are identical. For rod-like shapes, a flow rate decrease was detected with increasing aspect ratio comparing 2 samples of glass rods with 2 samples of plastic rods \cite{ashourSM2017}.

In the present work, we investigate experimentally the effect of particle shape on the flow rate of a granulate out of a container.
We use custom-made particles which differ only in shape, while their volume and all other parameters are identical.
We also test the resistance of our samples against quasistatic shearing. Our experiments are complemented with discrete element modelling (DEM).

\begin{figure}[!htbp]
    \includegraphics[width=0.99\linewidth]{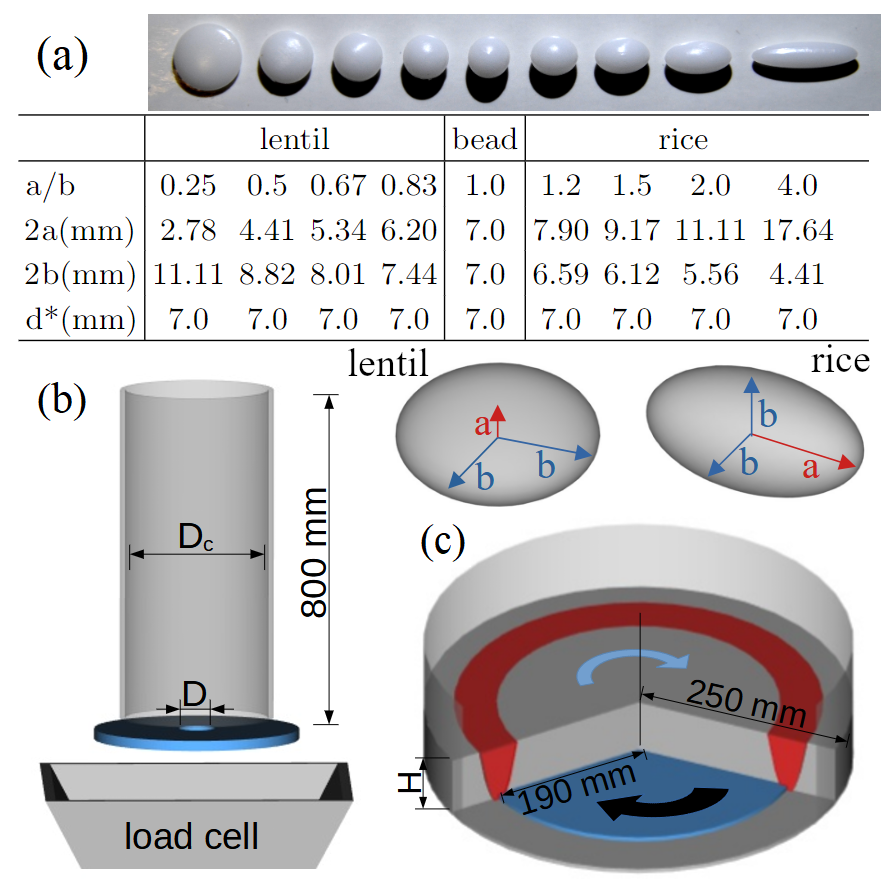}
    \caption{(a) Photographs and dimensions of the particles. All particles are rotational ellipsoids, they have the same volume, equivalent to the volume of a sphere with the diameter $d^*=7$ mm.  (b)-(c) Schematic diagrams of the experimental setups: silo and cylindrical split-bottom shear cell.}
    \label{materials-setups}
\end{figure}

In the experiments, we used 9 different samples of Polyoxymethylene (POM) rotational ellipsoids (produced by injection moulding by Yuyao Strong Co., China \cite{PARTICLES}).  Each sample contained 50000 identical particles, see Fig.~\ref{materials-setups}(a) for photographs of the particles and their characteristic dimensions. The flow rate experiments were performed using an acrylic cylinder with an inner diameter of either $D_\text{c}=172$ mm or 144 mm and a length of 800 mm with an orifice at the bottom with adjustable diameter $D$ (see Fig.~\ref{materials-setups}(b)). The granulate was filled into the cylinder manually, and after opening the orifice, we recorded the flow rate by measuring the weight of the discharged mass with a load cell. In a complementary experiment, we measured the resistance of the granulate against shearing in a cylindrical split-bottom shear cell (see Fig.~\ref{materials-setups}(c)). Here, the middle part of the sample was rotated with a rotating plate under it and thereby stationary shear was applied in the shear zone (see red region in Fig.~\ref{materials-setups}(c))  between the moving and standing regions. The applied torque was measured during stationary shearing.

The DEM implementation handles non-spherical  particles and their contact interaction using the superquadric equation \cite{Lu2015}. In particular, we employed a self-written GPU-NVIDIA implementation as a  parallelization procedure  \cite{Pongo2023} based on  \cite{podlozhnyuk2017efficient,navarro2013determination}, which allowed the examination  of system sizes comparable with the experiments.  
A superquadric is defined by the length of its half-axes $a$, $b$, $c$, and the blockiness parameters $n_1$ and $n_2$. To mimic the experiments, the parameters $n_1 = n_2 = 2$ were set, representing ellipsoids. Moreover, the values of $a$, $b \equiv c$ were the same as those of the experimental particles.
Complementarily, additional simulations were also done with triaxial ellipsoids ($a \neq b \neq c$) to check the generality of the results \cite{supp}.
In all the cases, the system is composed of $N=50000$ mono-disperse superquadrics, of the same equivalent diameter $d^*$.
POM has relatively low surface friction, therefore in the simulations presented here we use an inter-particle friction coefficient of $\mu=0.3$ \cite{laursenWEAR2009}.
After defining the particle-particle and particle-wall contact forces, the DEM computes the movement of each particle, see more details as supplemental material \cite{supp}. 
For the computation of the density, velocity and stress macroscopic fields, we have taken advantage of a useful coarse-graining technique, described in \cite{Babic1997,Goldhirsch2010,Weinhart2013,artoniPRE2015}. 

%
\begin{figure*}[!hbtp]
 \centering
  \includegraphics[width=1.0\textwidth]{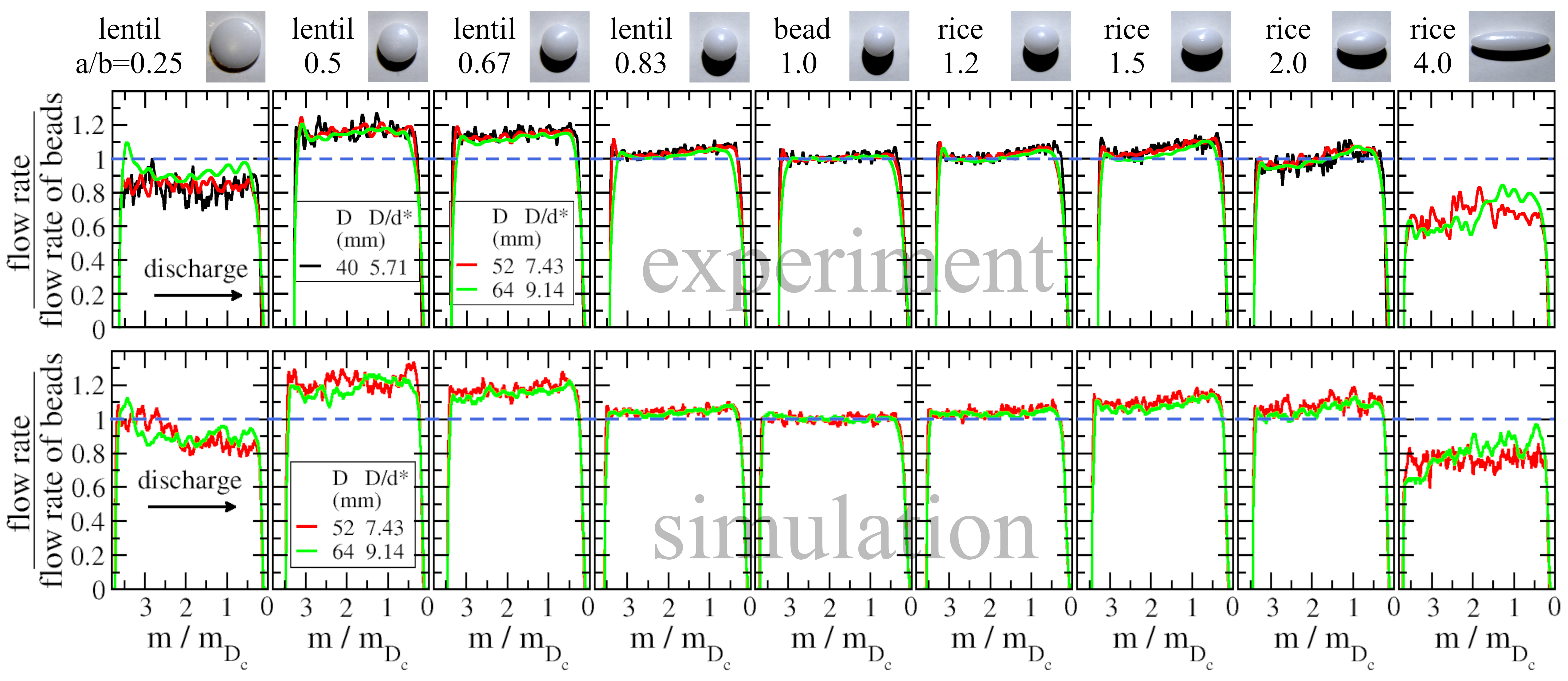}
    \caption{Normalized flow rate as a function of the mass in the cylinder for all 9 ellipsoidal samples during the discharge process. The average flow rate for beads is calculated in the range of $1.4<m/m_{\text{D}_{\text{c}}}<2.8$. Top row: experimental data obtained in the cylinder with $D_\text{c}=172$ mm with orifice sizes of $D=40$, 52 and 64 mm. Each curve corresponds to the average of 5 measurements. Bottom row: numerical results for the same setting. The curves correspond to a single run except for the cases of $a/b=0.25$ and $a/b=4.0$ for which 4 runs are averaged.}
 \label{flow-curves}
\end{figure*}

The evolution of the normalized flow rate obtained in both experiments and 
simulations are presented as a function of the mass in the cylinder during the discharge process in Fig.~\ref{flow-curves}.
The mass in the cylinder is normalized by the mass corresponding to the filling height of $D_\text{c}$. The data-sets are normalized by the average flow rate of beads. The average flow rate is calculated using data in the middle of the discharge process, i.e. in the range of $1.4<m/m_{\text{D}_{\text{c}}}<2.8$. Similar plots were obtained in the smaller silo ($D_\text{c}=144$ mm).
We mention that for a narrow cylinder, a surge is observed at the end of the discharge process for a certain range of the orifice diameter \cite{pongoNJP2022}. For most of our current measurements, there was no surge, for those where a surge occurs, we average the flow rate before the surge.

\begin{figure*}[!htbp]
 \centering
 \includegraphics[width=0.95\textwidth]{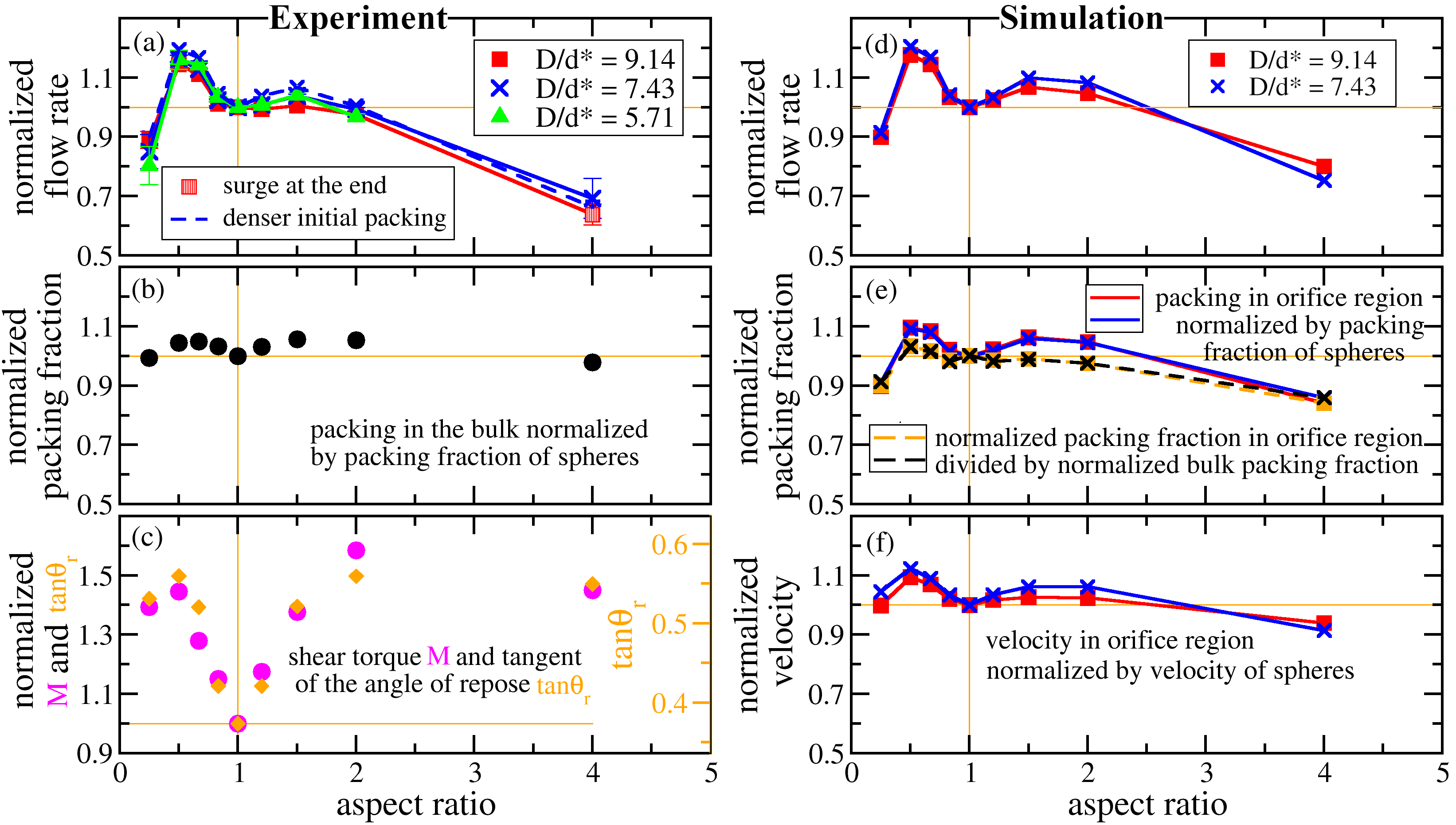}
   \caption{(a-c) Experimental results: (a) average flow rate (b) static packing fraction and (c) shear torque and tangent of the angle of repose as a function of particle aspect ratio. All quantities normalized by the value obtained for beads. The average flow rate was obtained by averaging the data in the range of $1.4<m/m_{\text{D}_{\text{c}}}<2.8$. (d-f) Numerical results: (d) average flow rate, (e) packing fraction and (f) velocity as a function of aspect ratio.}
 \label{flow-rate}
\end{figure*}

We summarize our findings by plotting the average flow rate as a function of the particle aspect ratio in Figs.~\ref{flow-rate}(a),(d). Starting from the spherical shape and going toward lentil-like shapes we first find a clear increase of the flow rate having a maximum around the aspect ratio of $a/b\approx 0.6$ and then a strong decrease for the grains with $a/b=0.25$. When we go in the direction of elongated grains, around the aspect ratio of about $a/b\approx 1.5$ we again find a peak, but this is smaller than  for the case of lentils, and the effect is more pronounced for the DEM data than in experiments (compare Figs.~\ref{flow-rate}(a)  and \ref{flow-rate}(d)). Above the aspect ratio of $a/b=2$ the flow rate clearly decreases. In Fig. \ref{flow-rate}(a) we also included a curve obtained in experiments with denser initial packing, which shows the same trend as the other curves. A recent work has highlighted the importance of the shape of the orifice rim on the flow rate \cite{wiacekSR2023}, thus we performed further experiments with a conical orifice that also show the same trend. Details about the measurements with denser initial packing and a conical orifice can be found in the supplemental material \cite{supp}. 
We also mention that the above results are consistent with other experimental observations in a conical hopper, where the discharge rate is found to be larger for lentil-like grains ($a/b=0.6$) and slightly lower for rice-like grains ($a/b=3$) compared to spheres \cite{Hanif2023}.

In order to explore the nonmonotonic nature of the flow rate curves (Figs.~\ref{flow-rate}(a),(d)), we first investigate whether this can be rather related to variations in packing fraction or grain velocities through the orifice. 

In the {\it experiments}, we can measure the bulk packing fraction of the initial state of each sample by measuring the weight of the sample in dry state and then submerged in water. As we see in Fig.~\ref{flow-rate}(b) the initial packing in the cylinder slightly depends on the grain shape with a nonmonotonic dependence for both lentil-like and ricelike shapes. The densest packings correspond to aspect ratios $a/b=0.5$ and $a/b=2$, and are about $6\%$ denser than the packing of spheres. A very similar dataset (not shown here) is obtained from our DEM simulations. All this is consistent with the results of earlier numerical calculations predicting similar shape dependence for the random close packed density \cite{donevSCI2004} and poured density \cite{ganPT2018} for both elongated and flat ellipsoids. Altogether, for the case of lentils the modulation of the initial bulk packing fraction with shape is small compared to the changes in flow rate, so it does not fully explain the flow rate behavior (Figs.~\ref{flow-rate}(a),(d)). 

\begin{figure}[htbp!]
    \centering
    \includegraphics[width=1.0\linewidth]{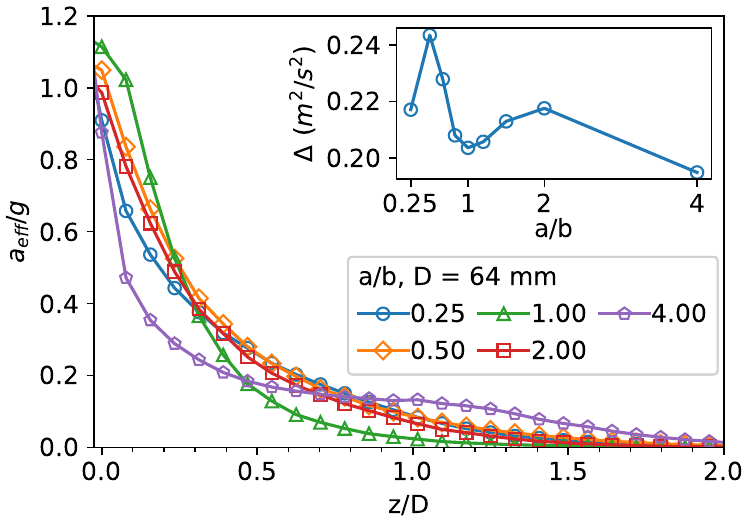}
       \caption{Average acceleration of grains in the cylindrical region of diameter $D$ above the orifice normalized by gravity as a function of the normalized distance $z/D$ from the orifice. The inset shows $\Delta = v^2_{\rm exit}/2$ at the orifice as a function of the aspect ratio.}
    \label{height-velo}
\end{figure}

In the {\it numerical simulations}, we can quantify the packing fraction as well as the velocity of the grains in the orifice region (see (Figs.~\ref{flow-rate}(e),(f)). As we see, they both show nonmonotonic shape dependence, and for the friction coefficient used here ($\mu=0.3$) they give approximately an equal contribution to form the trend observed in the flow rate. We mention that for $\mu=0.5$ the contribution of grain velocity becomes stronger (see supplemental material). Analyzing the data of the packing fraction first (Fig.~\ref{flow-rate}(e)), we find that dividing the normalized packing fraction in the orifice region by the normalized bulk packing fraction the resulting curve becomes nearly shape independent in the range of $0.5\leq a/b\leq 2$.
This means, that the flow does not significantly impact the packing in the orifice region for slightly elongated or flat ellipsoids, so the two peaks on the packing fraction curves basically come from the peaks in the bulk packing fraction.  
Focusing on the nonmonotonic trend of the particle velocity, we plot the average acceleration above the orifice as a function of the distance from the orifice in Fig.~\ref{height-velo}. Far from the orifice, the velocity and acceleration of all types of grains are negligible.
Interestingly, grains with non-spherical shape start accelerating from a higher position than the spheres, but as they get close to the orifice their acceleration becomes slightly smaller (compared to the case of spheres). Recovering the analysis done in \cite{rubiolargoPRL2015}, the vertical acceleration profile reads as $a_{\text{eff}}(z) = dv_z/dt=v_z \, dv_z/dz$.
Thus, a test particle coming downward in the center, far from the orifice, will gain $v_{\rm exit}$ vertical velocity arriving at the orifice level, based on the integral $\Delta = \int_{0}^{\infty} a_{\text{eff}}(z)dz = v_{\rm exit}^2/2$.
The inset of Fig.~\ref{height-velo} illustrates  $\Delta$ as a function of particle aspect ratio $a/b$. Remarkably, it resembles the nonmonotonic dependence of the vertical velocity appearing in Fig.~\ref{flow-rate}(f). Additionally, we also looked into the details of particle orientations in the orifice region. The analysis indicates that the non-sphericity significantly impacts the particle orientation and ordering with respect to flow (for details, see the supplemental material \cite{supp}). Interestingly, we find that for lentils and for short rice grains the flow rate and apparent particle size anti-correlate, i.e.~particles with a smaller cross section towards the orifice display a higher flow rate. 

\setlength{\parskip}{0pt}

Finally, coming back to the experimental data, we analyze how the resistance of the material against shearing is changing with grain shape, and whether this correlates with the flow rate behavior.
We quantify this by measuring the shear torque needed for a stationary quasi-static rotation of the sample in the split-bottom shear cell.
The normalized shear torque is presented as a function of the aspect ratio in Fig.~\ref{flow-rate}(c). The data were obtained using samples with the volume of 7 liters corresponding to a filling height of $H\approx3.5$ cm. The data were normalized first by the weight of the sample and then by the normalized shear torque obtained for  beads. As we see both elongated and flat shapes are characterized by an increased shear resistance compared to the case of spheres. The increase in the shear torque reaches about $50\%$ for both rice-like and  lentil-like ellipsoids, with aspect ratios of $a/b=2$ and 0.5, respectively. The increasing tendency is in accordance with our previous numerical findings on elongated shapes in simple shear \cite{nagyNJP2020} and inclined plane flow \cite{hidalgoPRF2018}, as well as with the tangent of the angle of repose ($\tan{\theta_\text{r}}$), which is also presented in Fig.~\ref{flow-rate}(c). The data for $\tan{\theta_\text{r}}$ were obtained in a quasi-two-dimensional cell \cite{supp,fanPRE2012} as an other measure to characterize the frictional property of the granulate. 
We emphasize that the data in Fig.~\ref{flow-rate}(c) evidences larger bulk friction for both elongated and flat grains compared to the case of spherical particles,  which would suggest smaller flow rate through a constriction for such particles. 
This is not in line with our observations on the flow rate, as described above.
Thus, the trend in the flow rate can not be explained by the shape dependence of the shear resistance and that of the angle of repose.

In summary, we find a surprising nonmonotonic behaviour of the silo flow rate as a function of grain shape for rotational and triaxial ellipsoids. Slightly non-spherical particles discharge faster than spheres, the effect is stronger for lentil-like shapes than for rice-like shapes. Analyzing the packing fraction and grain velocity in the orifice region, we find that they both contribute
to the nonmonotonic tendency of the flow rate. The contribution of the packing fraction mainly originates from the nonmonotonic shape dependence of the bulk packing fraction, while the contribution of grain velocity is related to two factors: (i) the height above the orifice line at which grains start accelerating (this is larger for non-spherical grains than for spheres), and (ii) the rate at which they speed up, which certainly depends on the dissipation in the region right above the orifice. The optimum of these two effects is found for the slightly flat lentil-like shapes which produce the largest discharge rate in the silo.
Finally, the resistance of the granulate against shearing considerably increases for both lentil-like and rice-like particles (compared to spheres), thus its trend does not coincide with the trend of the flow rate.

\begin{acknowledgments}
The authors acknowledge financial support from the European Union by Horizon 2020 Marie Sk\l{}odowska-Curie grant ''CALIPER'' (No. 812638) and COST action ''ON-DEM'' (No. CA22132). We are thankful for the discussions with J.~A.~Dijksman and J.~van der Gucht and to V.~Kenderesi for technical assistance.  
RCH acknowledges the Ministerio de Ciencia e Innovación (Spanish Government) Grant PID2020-114839GB-I00 funded by MCIN/AEI/10.13039/501100011033. TP acknowledges support by the Startup Grant from DKU. The Collective Dynamics Lab is partly sponsored by a philanthropic gift. BF and TP should be considered co-first authors.
\end{acknowledgments}

\bibliography{ms}

\end{document}


\title{Supplemental material: Effect of particle shape on the flow of an hourglass}
\author{Bo Fan,\textit{$^{1,3}$} Tivadar Pong\'o,\textit{$^{2,4,1}$}
Ra\'ul Cruz Hidalgo,\textit{$^{2}$} and Tam\'as B\"orzs\"onyi\textit{$^{1}$} 
} 

\affiliation{
$^1$Institute for Solid State Physics and Optics, HUN-REN Wigner Research Centre for
Physics, P.O. Box 49, H-1525 Budapest, Hungary\\
$^2$F\'isica y Matem\'atica Aplicada, Facultad de  Ciencias, Universidad de Navarra, Pamplona, Spain\\
$^3$Physical Chemistry and Soft Matter, Wageningen University $\&$ Research, Wageningen, The Netherlands \\
$^4$Collective Dynamics Lab, Division of Natural and Applied Sciences, Duke Kunshan University, 215306, Kunshan, Jiangsu, China}

\maketitle
 \section*{(A) Experiments with denser initial packing and a conical orifice}

Additional measurements were performed to investigate the effect of the initial packing fraction as well as the shape of the orifice. 

\begin{figure}[htbp]
\centering
\includegraphics[width=1.0\columnwidth]{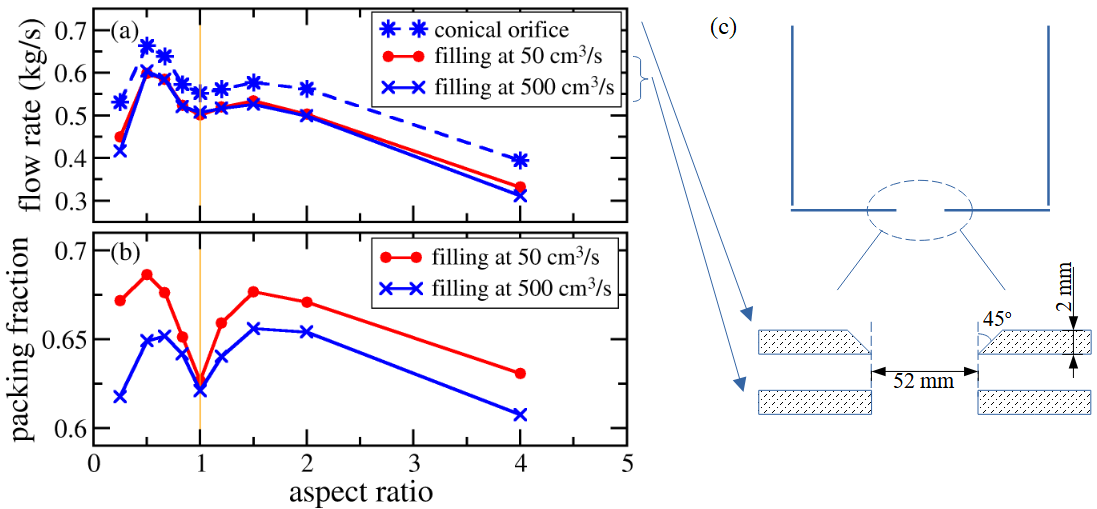}
\caption{Flow rate (a) and packing fraction (b) as a function of the grain aspect ratio. The data are in physical units (not normalized). The curves correspond to (x) normal initial packing, i.e. the same measurement as in Fig. 3 of the manuscript with normal filling procedure, ($\bullet$) denser initial packing obtained with a slow filling of the silo, and ($*$) using a conical orifice with a half cone angle of 45$^\circ$.  Orifice size: $D/d^*=7.43$ for all three measurements.
}
\label{fig1}
\end{figure}

In the first set of measurements, we prepared the initial condition with a higher packing fraction. This was achieved by pouring the grains 10 times slower (about $50$ cm$^3$/s instead of $500$ cm$^3$/s). This procedure led to better ordering of the grains, most of them lying horizontally. In Figure \ref{fig1}(a)-(b), we plot the silo flow rate and the packing fraction of the material in the silo in physical units (not normalized) as a function of the grain aspect ratio. As we see, the flow rate curve is essentially the same for the system with denser initial packing (filling at $50$ cm$^3$/s) as for the less ordered slightly less dense initial packing (filling at $500$ cm$^3$/s). The density increase due to filling at $50$ cm$^3$/s (compared to filling at $500$ cm$^3$/s) is in between 1-9$\%$, depending on grain aspect ratio, with the strongest density increase for lentils with aspect ratio 0.25, and the smallest density increase for spheres (see Fig. \ref{fig1}(b)). Thus we can conclude, that such variations in the initial packing do not change the non-monotonic trend of the flow rate as a function of grain aspect ratio, moreover the actual values of the flow rate remain essentially the same.

In the second set of measurements, we used a conical orifice with a half-cone angle of 45$^\circ$. See  Fig. \ref{fig1}(c) for the sketch of both types of orifices (conical and simple cylindrical).
As we see in Fig. \ref{fig1}(a), such an orifice leads to a different (slightly higher) flow rate compared to the case of the simple cylindrical orifice, in agreement with other studies \cite{wiacekSR2023}, but notably, the non-monotonic trend of the flow rate with grain aspect ratio is preserved.

\begin{figure}[htbp]
\centering
\includegraphics[width=1.0\columnwidth]{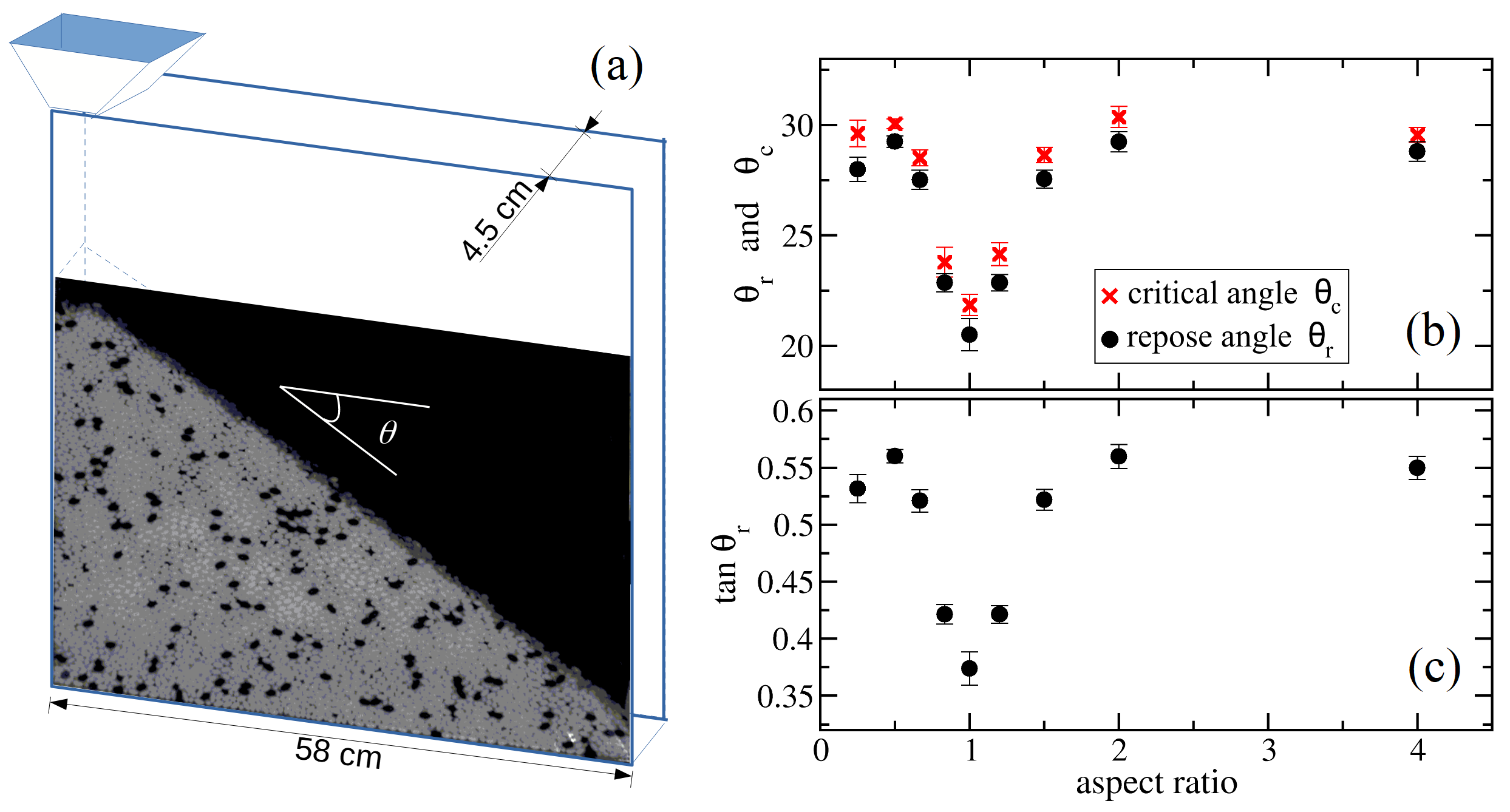}
\caption{\TB{(a) Sketch of the quasi-two-dimensional cell used for the characterization of the slope angle of a pile. (b) The critical angle $\theta_\text{c}$ and the repose angle $\theta_\text{r}$ as a function of the grain aspect ratio. (c) The tangent of the repose angle $\tan{\theta_\text{r}}$ as a function of the grain aspect ratio.}
}
\label{repose}
\end{figure}

\TB{\section*{(B) Characterization of the angle of repose}

We used a quasi-two-dimensional cell to characterize the slope angle of a pile as shown in Fig.~\ref{repose}(a) \cite{fanPRE2012}. In this experiment, grains were slowly added to the pile through the hopper indicated on the left side of the image. This leads to avalanches propagating down on the surface of the pile intermittently. The process was recorded, and the resulting movies were analyzed using digital image processing. The critical angle $\theta_\text{c}$ was measured before the avalanche started, and the repose angle $\theta_\text{r}$ was measured after the avalanche ceased and the surface of the pile was again at rest. A dozen of avalanches were measured for each material, and the average values of $\theta_\text{c}$ and $\theta_\text{r}$ are shown in Fig.~\ref{repose}(b) as a function of grain aspect ratio. The tangent of the repose angle $\tan{\theta_\text{r}}$ $-$ which is a widely used parameter to characterize the frictional property of a granulate $-$ is presented in Fig.~\ref{repose}(c) as a function of the grain aspect ratio.}

\section*{(C) Discrete element implementation}

The contact detection procedure is applied to all contacting particles $i$ and $j$, and it provides their plane of contact, described by the contact point and a normal vector $\hat{n}$, and the overlap distance $\delta$. We use a linear spring-dashpot model to compute the interaction force  $\vec{F}_{ij}$. It includes a  normal ($F^n_{ij}$) and a tangential ($F^t_{ij}$) contribution.  The normal component is $F^n_{ij} = -k_n \delta - \gamma_n v^n_{rel}$ where $k_n$ is the spring constant,  $\gamma_n$ is the damping coefficient, and $v^n_{rel}$ represents the normal relative velocity between the particles. 
The tangential component is $F^t_{ij} = -k_t \xi - \gamma_t v^t_{rel}$, where $k_t$ is the spring constant, $\gamma_t$ is the damping coefficient and $\xi$ and $v^t_{rel}$ are the tangential relative displacement and relative velocity of the particles.

In all the calculations, the particle density was $\rho_p = 1410 \,$kg/m$^3$, and the stiffness was set to $k_n = 2 \cdot 10^5 \,mg / d^*$ and  $k_t = 2/7 \cdot k_n$. 
The particle friction was $\mu = 0.3$ and the restitution coefficient was $e_n = 0.9$, which yielded the dissipation coefficients $\gamma_n \equiv \gamma_t$ using the formula in \cite{navarro2013determination}. \TB{Additional simulations with larger particle friction coefficients (0.5 and 0.7) resulted in similar non-monotonic curves for the flow rate as a function of grain aspect ratio.}

 \section*{(D) Numerical simulations with higher friction coefficient and non-rotational ellipsoids}

  \vspace{-0.3cm}
\TB{In order to check the generality of our observations, further numerical simulations were carried out with particles with a higher friction coefficient ($\mu=0.5$), as well as with} ellipsoidal particles of three different axis lengths. 

\begin{figure}[htbp]
    \centering
    \includegraphics[width=0.85\columnwidth]{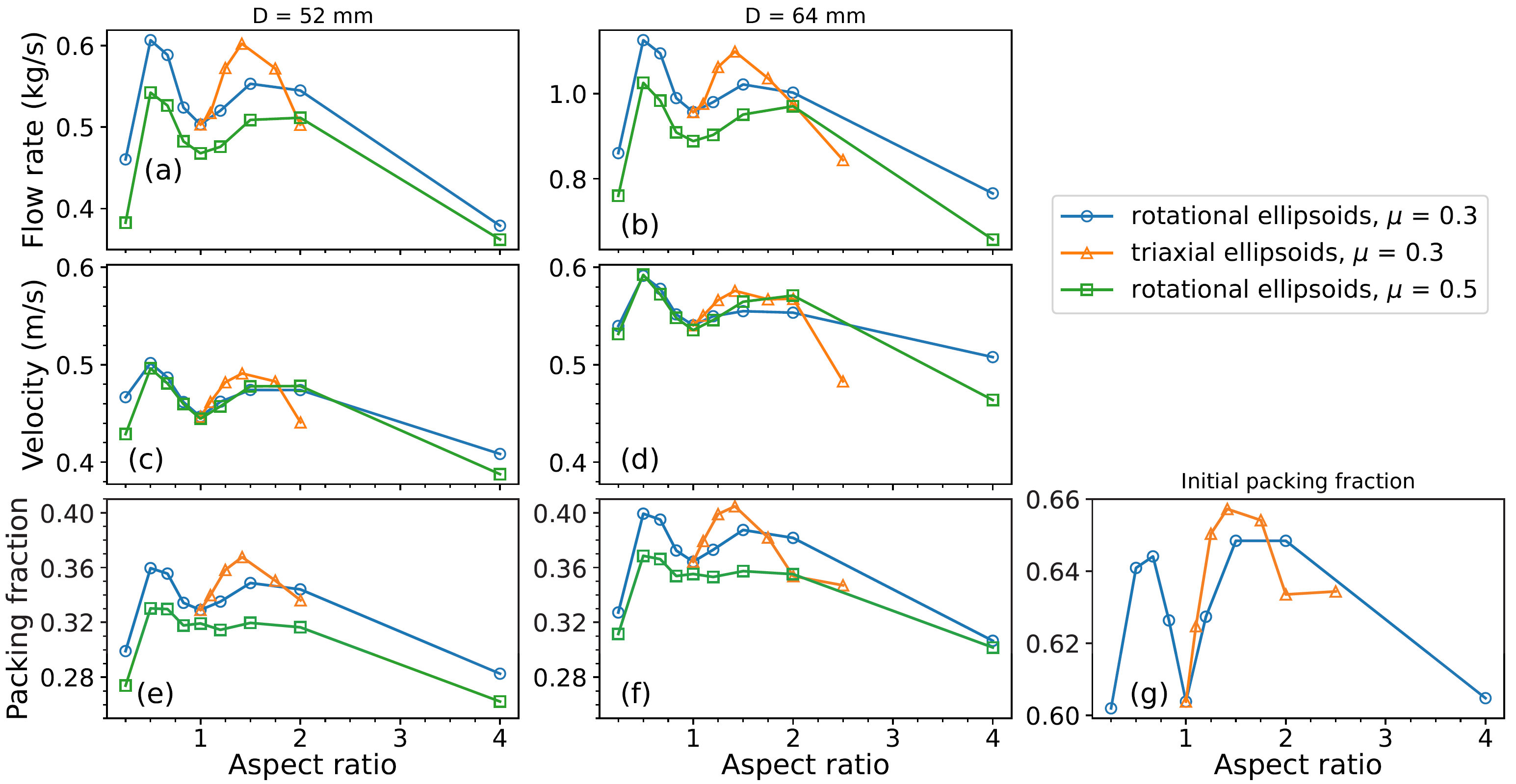}
    \vspace{-0.35cm}
    \caption{Flow rate (a-b), velocity (c-d) and packing fraction (e-f) at the orifice region as a function of the particle aspect ratio for triaxial (\textcolor{orange}{$\vartriangle$}) and rotational ellipsoids (\textcolor{blue}{$\circ$}) for comparison. The first column corresponds to an orifice size of $D = 52$ mm, the second to $D = 64$ mm. The overall packing fraction in the silo before the discharge is shown in panel (g).}
    \label{fig:flowrate_phi_vel_vs_aspratio_triaxial}
\end{figure}

\TB{In Fig.~\ref{fig:flowrate_phi_vel_vs_aspratio_triaxial} we see that the non-monotonic trend of the flow rate with particle aspect ratio prevails for particles with larger friction, and $-$ as expected $-$ we find a slightly lower flow rate for increased inter-particle friction.}

\vspace{-0.299cm}

\vspace{-0.3cm}
Since triaxial ellipsoids have three different axes, there are two free parameters that can be set to create an ellipsoid with a fixed volume. Here we used particles that can be described by one aspect ratio $\alpha$ such that $a/b = \alpha;\ c/b=\alpha^{-1}$ following the definition in Donev et al.~\cite{donevSCI2004}. Note that this way the ratio of the longest to shortest axis is $\alpha^2$.

\begin{figure}[htbp]
    \centering
    \includegraphics[width=0.99\columnwidth]{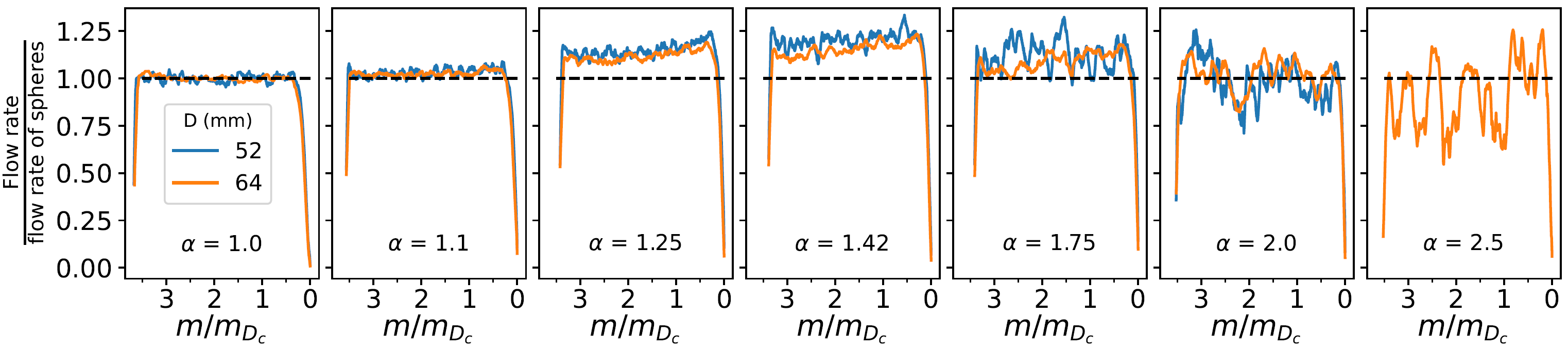}
   \vspace{-0.35cm}
    \caption{Normalized flow rate for triaxial ellipsoids ($\alpha=a/b=b/c$) as a function of the mass in the cylinder during the discharge. The different subplots correspond to the different aspect ratios $\alpha$ which is indicated above. The curve for $\alpha=2.5$, $D = 52$ mm is missing because the system clogged early during the flow. }
    \label{fig:flowrate_vs_mass_triaxial_ellipsoids}
\end{figure}

In Fig.~\ref{fig:flowrate_vs_mass_triaxial_ellipsoids} the flow rate normalized by the average flow rate of the spheres is plotted as a function of the mass in the silo. The x axis is scaled by $m_{D_c}$, which is the mass that is in the silo that is filled to a height of $D_c$. We average the flow rate of the different aspect ratio triaxial particles using data in the range of $1.4 < m/m_{D_c} < 2.8$. These are displayed in Fig.~\ref{fig:flowrate_phi_vel_vs_aspratio_triaxial}(a),(b) with the orange triangle symbols alongside the average flow rates for the rotational ellipsoids (blue circles). The discharge rate first increases for slightly nonspherical particles and reaches a maximum at around $\alpha=1.4$, after which it decreases with increasing $\alpha$.

\vspace{-0.27cm}

\vspace{-0.3cm}

In order to see whether the particle packing density at the orifice or their velocity causes this non-monotonic trend, we average these two coarse-grained fields at the region of the orifice.
The results plotted in Fig.~\ref{fig:flowrate_phi_vel_vs_aspratio_triaxial}(c-f) show that both the packing fraction and the velocity of particles exiting the silo exhibit the trend of the flow rate. Additionally, we show the initial packing fraction of these triaxial particles in Fig.~\ref{fig:flowrate_phi_vel_vs_aspratio_triaxial}(g).
Altogether, the triaxial ellipsoids manifest a similar non-monotonic dependence of the flow rate on the aspect ratio, which adds to the weight and generality of our findings.

\newpage

\section*{(E) Particle orientations in the region of the orifice}

\begin{figure}[htbp]
    \centering
    \includegraphics[width=0.7\columnwidth]{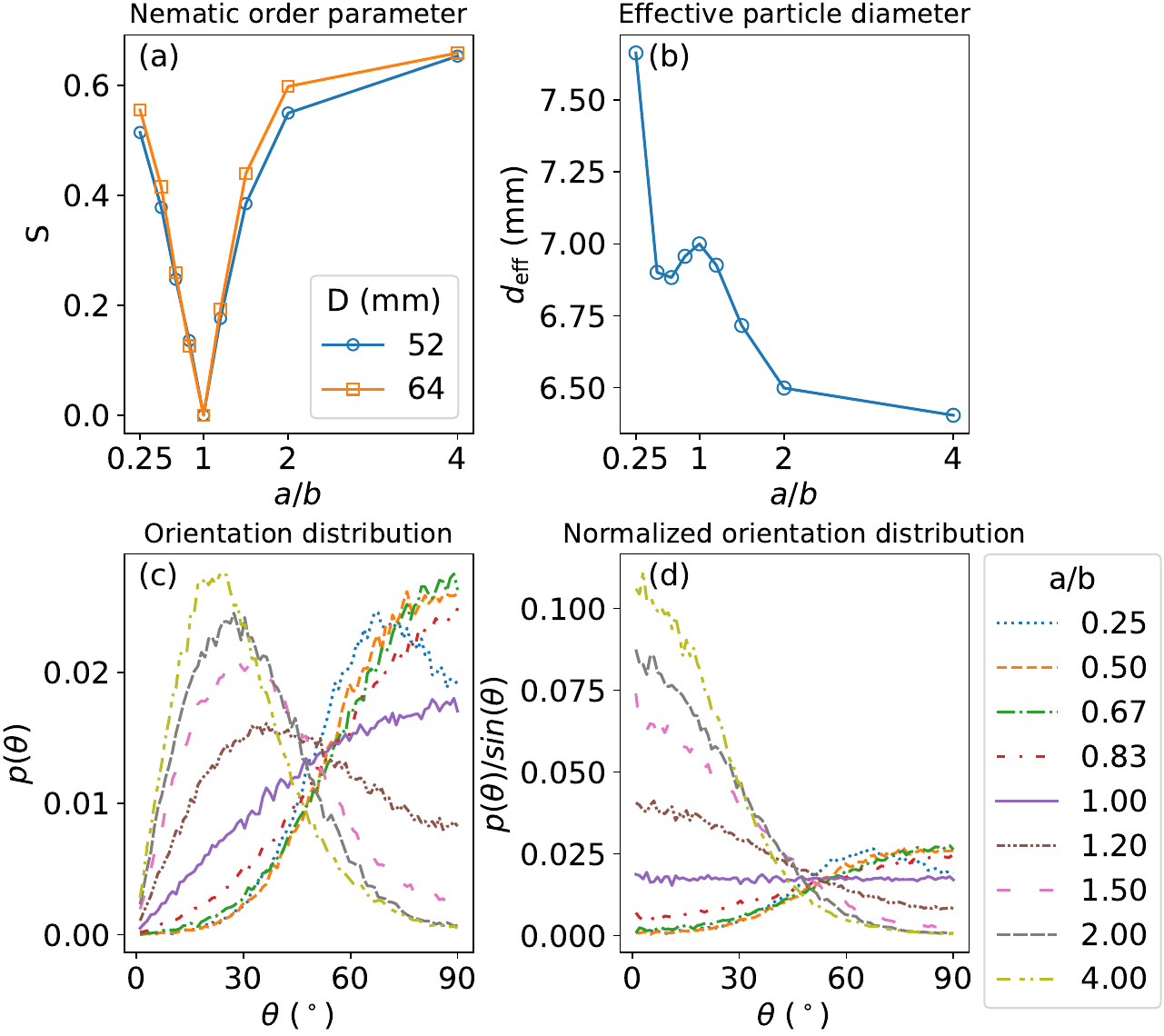}
    \caption{(a) Order parameter and (b) effective grain diameter as a function of the grain aspect ratio $a/b$. (c) Distribution of the particle orientations $p(\theta)$, where $\theta$ is measured with respect to the vertical axis, (d) $p(\theta)$ normalized by the orientation distribution of a system with random particle orientations.}
    \label{order-orientation}
\end{figure}

Next, we analyze the orientation of particles and look for an eventual correlation between particle alignment and flow rate. As we described in earlier work, the shear flow leads to orientational ordering of the grains 
\cite{borzsonyiPRL2012,borzsonyiPRE2012,borzsonyiNJP2016}, which we quantify by a nematic order parameter $S$. This is calculated by diagonalizing the symmetric traceless order tensor \T{}:
\begin{equation}
T_{ij}= \frac{3}{2N} \sum\limits_{n=1}^N \left[{\ell}^{(n)}_i
{\ell}^{(n)}_j
-\frac{1}{3} \delta_{ij}
\right] \quad ,
\end{equation}
where $\vec {\ell}^{(n)}$ is a unit vector along the axis of particle
$n$, and the sum is over all $N$ particles present in a certain volume element. 
The largest eigenvalue of \T{} is the primary nematic order parameter $S$ \cite{borzsonyiPRL2012,borzsonyiPRE2012}.
Random grain orientations would lead to $S=0$, while $S=1$ corresponds to a perfect alignment with all grains parallel to each other. Fig. \ref{order-orientation}(a) shows the order parameter as a function of grain aspect ratio $a/b$ in the orifice region. As we see, nematic ordering quickly increases as we go from spheres towards rice-like or lentil-like grains. We can define an effective grain diameter $d_{\rm eff}=2\sqrt{A/\pi}$ where $A$ is the mean cross section of the particles as projected on the orifice plane. As we see in Fig.~\ref{order-orientation}(b), for lentils, we get an interesting non-monotonic trend of  $d_\text{eff}$ as a function of $a/b$, which nicely anticorrelates with the flow rate, i.e.~we observe larger flow rate for particles with smaller cross-section. At the same time $d_\text{eff}$ is monotonically decreasing with $a/b$ for rice like grains.

 Looking into the details of particle orientations in the orifice region we plot the orientation distributions $p(\theta)$ in Fig.~\ref{order-orientation}(c), where $\theta$ is measured with respect to the vertical axis. These curves show the same tendency for rice and lentils: both types of grains get aligned towards the orifice with their smaller cross section. For rice the long axis of the grains encloses a small angle with vertical, while for lentils the short axis of the grains is typically horizontal. We note that in this representation the curve describing a randomly oriented system is $p(\theta)\propto \sin(\theta)$, as we see for the spherical particles. Therefore it is useful to plot the normalized orientation distributions $p(\theta)/\sin(\theta)$ (see Fig.~\ref{order-orientation}(d)) which shows the particle orientations with respect to a random system. As we see, for rice-like grains we see a clear increase in ordering with $a/b$, while for lentils the aspect ratio dependence is much weaker. This is the reason why we have a non-monotonic $d_\text{eff}(a/b)$ curve for lentils, but not for rice. Altogether, we emphasize that for lentil-like and for short rice-like grains we find an inverse correlation between flow rate and effective grain size (see Figs.~\ref{fig1}(a), \ref{fig:flowrate_phi_vel_vs_aspratio_triaxial}(a,b) and \ref{order-orientation}(b)), i.e. particles facing the orifice with a smaller cross section display a larger flow rate.

\bibliographystyle{apsrev4-1}
\bibliography{ms}